\documentclass[journal=jacs,manuscript=article]{achemso}
\usepackage[utf8]{inputenc}
\usepackage[T1]{fontenc}
\usepackage[version=3]{mhchem}
\usepackage{soul}
\usepackage{bm}
\usepackage{threeparttable}
\usepackage{color}


\author{Qi Yu}
\affiliation{Department of Chemistry, Yale University, New Haven, Connecticut 06520, U.S.A.}
\email{q.yu@yale.edu}

\title{Multidimensional quantum calculation of the infrared spectra under polaritonic vibrational strong and ultrastrong coupling}

\begin{document}
\clearpage

\begin{abstract}
Recent experiments and theory demonstrate that the the ground state properties and chemical reactivity of molecules can be modified inside an optical cavity. The vibrational strong or ultrastrong coupling results in the formation of vibrational polaritons which are usually observed through infrared spectra (IR). Here, we provide a theoretical framework to conduct multidimensional quantum simulations of the infrared spectra when the molecule is interacting with cavity modes. Taking single water molecule as an example, combing with accurate potential energy and dipole moment surfaces, our implemented cavity vibrational self-consistent field/virtual state configuration interaction (cVSCF/VCI) is shown to be able to provide quantitative predictions of the IR spectra when the molecule is inside or outside the cavity. The spectral signatures of resonance splittings and blue/red shift of certain bands are found to be highly related with the frequency and polarization direction of the cavity modes. Further analyses of the simulated spectra shows that polaritonic strong vibrational coupling greatly induce the coupling between molecule's vibrational modes, indicating the intramolecular vibrational energy transfer may be significantly accelerated by the cavity.

\end{abstract}

\flushbottom
\maketitle

\thispagestyle{empty}
\clearpage

\section*{Introduction}
Strong light-matter interactions between molecular system and the electromagnetic field of an optical cavity provide new opportunities for modifying chemical reactivities and selectivities.\cite{Ebbesen2016_acr,Nori2019,Hirai2020,Herrera2020_1,Xiang2021_jcp,Dunkelberger2022,Li2022} As a hybrid state involving both infrared cavity mode and molecule's vibrational mode, molecular vibrational polariton has been shown both experimentally and theoretically to have the possibilities of influencing chemical reaction rates and vibrational energy transfer.\cite{Hutchison2012,Thomas2019,Lather2019,Hutchison2020,Ebbsen2019,Xiang2020,Grafton2021,Li2021_angew,Li2020_pnas,Yuanzhou2019,Yuanzhou2019_acs,Feist2020,Herrera2016,Huo2021_natcom,Huo2021_jpcl,Ribeiro2018,Cao2021,Gu2020,Feist2017,Rubio2017} A typical signature of vibrational polaritons is the observation of Rabi splitting of a vibrational peak in the molecular infrared spectrum. The dimensionless ratio $\eta$ between Rabi splitting frequency $\Omega_R$ and molecule's vibrational frequency $\omega_0$, defined as $\eta=\Omega_R/(2\omega_0)$, can be used to indicate the light-matter coupling strength. The criteria of $\eta \ge 0.1$ becomes a simple marker for the transition from the vibrational strong coupling (VSC) to the vibrational ultrastrong coupling (V-USC) regime.\cite{Thomas2019,Hirai2020,Trung2020,Li2020_pnas} Experimental techniques such as linear infrared spectroscopy and nonlinear two-dimensional infrared spectroscopy (2D-IR) have been employed to investigate the vibrational polaritonic dynamics in these regimes.\cite{Thomas2016,Thomas2019,Lather2021,Ebbsen2019,Xiang2020,Xiang2018} To fully understand the experimental spectra and vibrational dynamics, reliable theoretical analyses are needed. 

A series of theoretical investigations have also been conducted focusing on the electronic ground state properties, IR spectra and also reaction dynamics.\cite{Cao2021,Li2021_angew,Li2020_pnas,Yuanzhou2019,Yuenzhou2020,Yuanzhou2019_acs,Fischer2021,Herrera2016,Rubio2017,Rubio_2017_jctc,Rubio2018_1,Huo2021_natcom,Huo2021_jpcl,Ribeiro2018,Ribeiro2018_jpcl} Many of these theoretical works simplified the complicated multi-dimensional molecular system into a reduced-dimensional (usually 1D) problem, where simple potential energy and dipole moment functions are used.\cite{Herrera2016,Huo2021_jpcl,Huo2021_natcom,Fischer2021,Cao2021} Although these studies provide fruitful insights of vibrational polariton's impact on infrared spectra and chemical reactivities, the important multi-dimensional anharmonicities and mode couplings in the molecular system are often ignored. Recently, realistic multi-dimensional potential energy and dipole moment models were applied for theoretical investigation of VSC or V-USC systems.\cite{Li2020_pnas,Li2021_angew} For example, Li and coworkers successfully simulated the IR spectra of liquid water within an optical cavity based on their cavity molecular dynamics interfacing with the q-TIP4P/F water force field. However, the limited accuracy of these conventional force fields may have significant impact on the simulation results. Another route to investigate molecule's properties within cavity is \emph{ab initio} cavity quantum electrodynamics (QED) approaches such as QED density functional theory (QEDFT)\cite{Rubio2014,Rubio2017} and QED coupled cluster theory (QED-CC)\cite{Koch2020}. The quantum dynamics simulations or long time classical simulations usually require extensive single point calculations and thus these ``on the fly'' \emph{ab initio} methods can be computationally expensive. 

Apart from the quality of potential energy and dipole moment surfaces (PES/DMS), the choice of reliable method for theoretical simulations is also important. Focusing on IR spectrum, classical MD is an efficient and general approach to obtain the spectra through Fourier transformation of the dipole-dipole correlation function. However the classical MD approach misses important quantum effects like zero-point energy effect and simulated spectrum can not provide detailed information of the origin of the spectral peaks, such as anharmonic mode couplings.\cite{Yu2019} Quantum approaches including vibrational second order perturbation theory (VPT2),\cite{Nielsen,Barone2005} vibrational self-consistent field/virtual state configuration interaction (VSCF/VCI)\cite{vscf78,vscf86,VSCF3,VCI} and multiconfiguration time-dependent Hartree (MCTDH) method\cite{mey90:73,man92:3199} provides predictions of vibrational state energies along with detailed spectral assignments. The interested readers are referred to the literature for details and assessment of these approaches.\cite{Bowman2008,Bowman2019}

In this letter, we provide the theoretical framework for conducting quantum VSCF/VCI simulations of IR spectra of molecular systems within an optical cavity, denoted as cVSCF/VCI. In conjunction with spectroscopically accurate molecular potential energy and dipole moment surfaces, we successfully simulated the vibrational spectra of single water molecule interacting with cavity modes under different cavity frequencies, light-matter coupling strengths and cavity polarization directions. The accuracy of VSCF/VCI scheme is confirmed by the excellent agreement between theoretical and experimental IR spectra of \ce{H2O} outside the cavity. We demonstrate that by tuning the cavity frequency and coupling strength, the spectral positions of vibrational polaritons can change significantly along with observed Rabi splitting values. We also confirm the fact that the generation of vibrational polaritons is also sensitive to the cavity polarization direction. From detailed analysed of spectral lines and observation of unexpected splitting features, our full dimensional quantum simulations show that the vibrational strong coupling and ultrastrong coupling open new possibilities of improving mode couplings within the molecule and may greatly accelerate the intramolecular vibrational energy transfer. Our calculation results can be viewed as benchmarks for other approximate methods that can be used on much larger systems.

We start by introducing the widely used Pauli-Fierz Hamiltonian\cite{Rubio_2017_jctc,Rubio2017,Rubio2018_1} for the cavity quantum electrodynamics (QED):
\begin{equation}
\hat{H}_{\text{QED}} = \hat{H}_{M} + \hat{H}_{C} + \hat{H}_{CM} 
\end{equation}
Here, we use the exact Watson Hamiltonian\cite{Watson77} for nonlinear molecule in mass-scaled rectilinear normal modes, \textbf{Q}:
\begin{equation}
\begin{aligned}
\hat{H}_{M}=\frac{1}{2}\sum_{\alpha \beta}(\hat{J}_{\alpha}-\hat{\pi}_{\alpha})\kappa_{\alpha \beta}(\hat{J}_{\beta}-\hat{\pi}_{\beta})-\frac{1}{2}\sum_{i}^{N}\frac{\partial^2}{\partial Q_i^2}-\frac{1}{8}\sum_{\alpha}\kappa_{\alpha \alpha}+ V(\textbf{Q})
\end{aligned}
\end{equation}
where $\alpha(\beta)$ represent the $x, y, z$ coordinates,
$\hat{J_{\alpha}}$ and $\hat{\pi}_{\alpha}$ are the components of the total and vibrational angular momenta respectively, $\kappa_{\alpha\beta}$ is the
inverse of effective moment of inertia, $N$ is the number of normal modes, and $V(\textbf{Q})$ is the potential energy of the molecular system.
The second term in right hand side of Equation (1), $\hat{H}_C$, is the cavity photon field Hamiltonian such that:
\begin{equation}
\begin{aligned}
\hat{H}_{C}=\sum_k^{N_c}\hbar\omega_k(\hat{a}_k^{\dagger}\hat{a}_k+\frac{1}{2})=\sum_k^{N_c}\frac{1}{2}(\hat{p}_k^2+\omega_k^2\hat{q}_k^2)\\
\end{aligned}
\end{equation}
where $N_c$ is the number of cavity modes, $\omega_k, \hat{a}_k^{\dagger}, \hat{a}_k, \hat{p}_k, \hat{q}_k$ are the frequency, photonic creation, annihilation, momentum and position operators for the cavity mode $k$ respectively, and 
\begin{equation*}
\begin{aligned}
\hat{q}_k=\sqrt{\frac{\hbar}{2\omega_k}}(\hat{a}_k^{\dagger}+\hat{a}_k), \hat{p}_k=i\sqrt{\frac{\hbar\omega_k}{2}}(\hat{a}_k^{\dagger}-\hat{a}_k)
\end{aligned}
\end{equation*}
The last term in Equation (1), $\hat{H}_{CM}$, is then light-matter interaction term written as :
\begin{equation}
\begin{aligned}
&\hat{H}_{CM}=\sum_k^{N_c} \omega_k\hat{\boldsymbol{\mu}}\cdot \textbf{A}_k(\hat{a}_k+\hat{a}_k^{\dagger})+\frac{\omega_k}{\hbar}(\hat{\boldsymbol{\mu}}\cdot \textbf{A}_k)^2\\
\end{aligned}
\end{equation}
where $\textbf{A}_k=\textbf{e}_k\sqrt{\hbar/(2\omega_c\epsilon_0\tilde{V})}$, $\epsilon_0$ is the permittivity, $\tilde{V}$ is the effective cavity volume, $\textbf{e}_k$ is the polarization vector of the field.
Equation (4) can be expressed using $\hat{p}_k, \hat{q}_k$ as follows
\begin{equation}
\begin{aligned}
\hat{H}_{CM}=\sum_k^{N_c} \omega_k \sqrt{\frac{1}{\epsilon_0 V}}\hat{q}_k(\hat{\boldsymbol{\mu}}\cdot\textbf{e}_k)+\frac{1}{2}\frac{1}{\epsilon_0\tilde{V}}(\hat{\boldsymbol{\mu}}\cdot\textbf{e}_k)^2
\end{aligned}
\end{equation}
The last component in Equation (5) is the dipole self-energy term which is usually included for vibrational polariton system.\cite{Fischer2021} Combing Equation (2), (3) and (5), we reach the final expression of the Pauli-Fierz Hamiltonian for molecular system:
\begin{equation}
\begin{aligned}
\hat{H}&=\hat{H}_{M}+\sum_k^{N_c}\Big[\frac{1}{2}\hat{p}_k^2+\frac{1}{2}\omega_k^2(\hat{q}_k+\frac{1}{\omega_k}\sqrt{\frac{1}{\epsilon_0\tilde{V}}}\hat{\boldsymbol{\mu}}\cdot\textbf{e}_k)^2\Big]\\
\end{aligned}
\end{equation}
Define the light-matter coupling factor, $g=\sqrt{\omega_k/(2\epsilon_0\tilde{V})}$, the total Hamiltonian becomes:
\begin{equation}
\begin{aligned}
\hat{H}&=\frac{1}{2}\sum_{\alpha \beta}(\hat{J}_{\alpha}-\hat{\pi}_{\alpha})\kappa_{\alpha \beta}(\hat{J}_{\beta}-\hat{\pi}_{\beta})-\frac{1}{2}\sum_{i}^{N}\frac{\partial^2}{\partial Q_i^2}-\frac{1}{8}\sum_{\alpha}\kappa_{\alpha \alpha}\\
&+\sum_k^{N_c}\frac{1}{2}\hat{p}_c^2+V(\textbf{Q})+\sum_k^{N_c}\frac{1}{2}\omega_k^2(\hat{q}_k+\sqrt{\frac{2}{\omega_k^3}}g\hat{\boldsymbol{\mu}}\cdot\textbf{e}_k)^2\\
\end{aligned}
\end{equation}
As seen from Equation 7, the effective potential energy of the molecule-cavity system is
\begin{equation}
    V^{\text{eff}}(\textbf{Q},\textbf{q})=V(\textbf{Q})+\sum_k^{N_c}\frac{1}{2}\omega_k^2(\hat{q}_k+\sqrt{\frac{2}{\omega_k^3}}g\hat{\boldsymbol{\mu}}\cdot\textbf{e}_k)^2
\end{equation}

Now we introduce the cVSCF/VCI approach for simulating the IR spectra of molecular systems in an optical cavity. Analogous to the conventional VSCF method, under mean-field approximation, we represent a general quantum state as an $N+N_c$ mode wavefunction with direct product form, 
\begin{equation}
\begin{aligned}
&\Phi(\textbf{Q},\textbf{q})=\prod_{i=1}^N\phi_i(Q_i)\prod_{i=1}^{N_c}\chi_i(q_i) \\
\end{aligned}
\end{equation}
The variational principle is then applied to each modal function, subject to the constraint that each has a unit norm. This leads to a series of SCF equations,
\begin{equation}
\begin{aligned}
&\Big[T_i+\big\langle \prod_{j\ne i}^N\phi_j(Q_j)\prod_{j=1}^{N_c}\chi_j(q_j)\big|V^{\text{eff}}(\textbf{Q},\textbf{q})+T_c\big|\prod_{j\ne i}^N\phi_j(Q_j)\prod_{j=1}^{N_c}\chi_j(q_j) \big\rangle \Big]\phi_i(Q_i)=\varepsilon_i\phi_i(Q_i)\\
&\Big[T_i^c+\big\langle \prod_{j=1}^N\phi_j(Q_j)\prod_{j\ne i}^{N_c}\chi_j(q_j)\big|V^{\text{eff}}(\textbf{Q},\textbf{q})\big|\prod_{j=1}^N\phi_j(Q_j)\prod_{j\ne i}^{N_c}\chi_j(q_j) \big\rangle \Big]\chi_i(q_i)=\varepsilon_i\chi_i(q_i)\\
\end{aligned}
\end{equation}
where $T_i$ is the kinetic energy operator of each mode in molecule, $T_c$ is the Coriolis coupling terms in molecule, and $T_i^c$ is the kinetic energy operator of each cavity mode. All these terms have been defined in Equation (2), (3) and (7). \\
The SCF equations in Equation (11) are then solved in the usual iterative manner until self-consistency. The final solutions are 1D wavefuntions for each mode, in the environment of effective potentials that depend on the quantums state of other modes. Analogous to the configuration interaction (CI) method in electronic structure theory, the ground and virtual states from VSCF solutions, $\Phi_m(\textbf{Q},\textbf{q})$, can be used as the CI basis for the rigorous expansion of the exact wave function. Thus the vibrational CI (VCI) wave function is written as
\begin{equation}
\begin{aligned}
\Psi(\textbf{Q},\textbf{q})=\sum_m C_m\Phi_m(\textbf{Q},\textbf{q})=\sum_mC_m\prod_{i=1}^N\phi_i^{m}(Q_i)\prod_{i=1}^{N_c}\chi_i^{m}(q_i) \\
\end{aligned}
\end{equation}
The expansion coefficients in this VCI formula can be obtained by diagonalizing the Hamiltonian matrix. With the final VCI wave functions, the general expression for the IR intensity for the transition between two quantum eigenstates is
\begin{equation}
\begin{aligned}
I_{f \leftarrow i}&= \frac{8\pi^3N_{A}}{3hc(4\pi\epsilon_0)}\nu_{f \leftarrow i}\sum_{\alpha=x,y,z}|R_{\alpha if}|^2(n_i-n_f)\\
R_{\alpha if} &=\int \psi_i(\textbf{Q},\textbf{q})\mu_{\alpha}(\textbf{Q})\psi_f(\textbf{Q},\textbf{q})d\textbf{Q}d\textbf{q}
\end{aligned}
\end{equation}
where $N_A$ is the Avogadro’s number, $h$ is the Planck constant, $c$ is the speed of light, $\epsilon_0$ is the vacuum permittivity, $\nu_{f \leftarrow i}$is the the transition frequency, and $\mu_(\textbf{Q})$ is the dipole moment of the molecule. The factor ($n_i$,$n_f$) is the difference in the fraction of molecules in the initial and final states, and when the temperature is low, this term tends to 1 for the transitions originating from the ground vibrational state. Similar to the conventional VSCF/VCI approach, the n-mode representation (nMR) of the effective potential $V^{\text{eff}}(\textbf{Q},\textbf{q})$ and dipole moment is used which enables efficient numerical quadrature calculation of the matrix elements of potential and dipole components. More details of the n-mode representation and the VSCF/VCI approach can be seen in the Supporting Information (SI) and a recent review.\cite{Bowman2019}

The quantum cavity-VSCF/VCI (cVSCF/VCI) approach for the molecular system in an optical cavity has been implemented in the software MULTIMODE.\cite{multimode,Multimode2} As mentioned above, to conduct robust and multidimensional calculations on real molecules, accurate potential energy and dipole moment surfaces are required. Here, for the system of a single \ce{H2O} molecule coupled to cavity mode, we employed spectroscopically accurate PES of \ce{H2O} developed by Partridge et al.\cite{PS} and the highly accurate DMS of \ce{H2O} developed by Lodi et al..\cite{h2odip} These PES and DMS are also interfaced to the modified version of MULTIMODE for cavity system calculations. In all the cVSCF/VCI calculations throughout the paper, 4-mode representation (4MR) of the effective potential and 3-mode representation (3MR) of the dipole moment were used. The VCI excitation space is (10,10,10,10) for single, double, triple and quadruple excitations respectively, resulting in sufficiently large VCI matrix. As shown in Figure 1(a), the cavity is placed along the z axis and the single \ce{H2O} molecule is put in the x-y plane to maximize the coupling with the field. For the cavity modes, they have the polarization direction in the x-y plane and we will explore how the role of polarization direction plays in the final IR spectra. Figure 1(b) shows the dipole moment curves along x, y, and z axes for the \ce{H2O} molecule's bending, symmetric stretch and asymmetric stretch modes. Since the water molecule in put in x-y plane, the z-component of the dipole is always zero. For the bending mode, its motion results in dramatic change of the x-component of the dipole while no change for the y component. Similar behavior appears in the motion of symmetric stretch mode, although the change of x-component of dipole is less significant than that for the bending motion. As to the asymmetric stretch, only y-component of dipole changes significantly and almost no change for the x component.

\begin{figure}[htbp!]
\begin{center}
\includegraphics[width=0.8\textwidth]{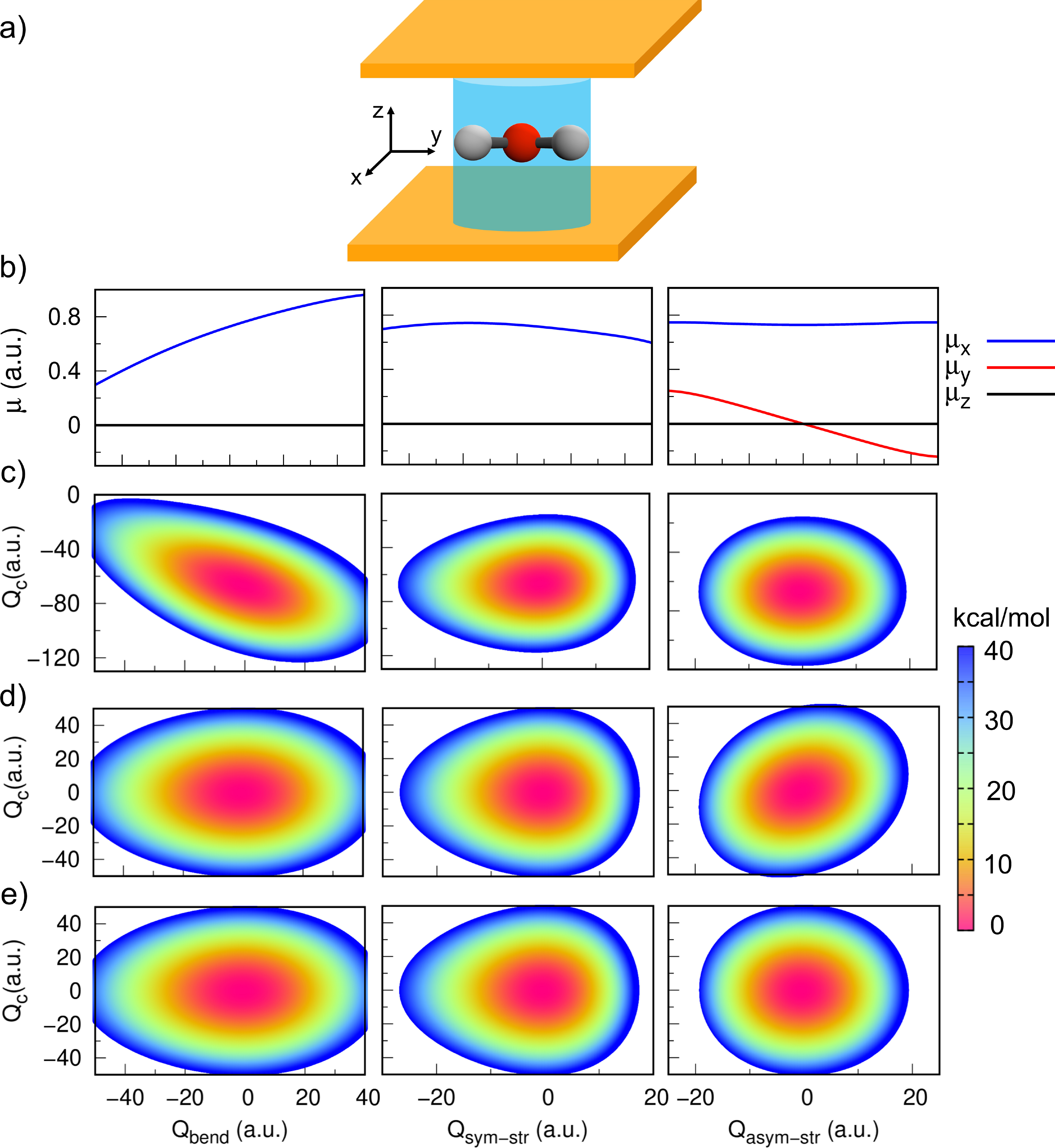}
\end{center}
\caption{(a) Scheme plot of a \ce{H2O} molecule placed inside an optical cavity where cavity direction is along z-axis and the \ce{H2O} molecule is put on x-y plane. (b) Dipole moment along \ce{H2O} bending (left), symmetric (middle) and asymmetric (right) stretches. (c)-(e) Cavity potential energy surface along cavity mode and different normal modes of \ce{H2O} where (c) molecule within the cavity with cavity mode's polarization direction along x axis, (d) molecule within the cavity with cavity mode's polarization direction along y axis, (e) molecule outside the cavity. The cavity mode frequency is 1594 cm$^{-1}$ and the light-matter coupling factor g is 0.04 a.u. for all cases.}
\end{figure}

\begin{figure}[htbp!]
\begin{center}
\includegraphics[width=0.5\textwidth]{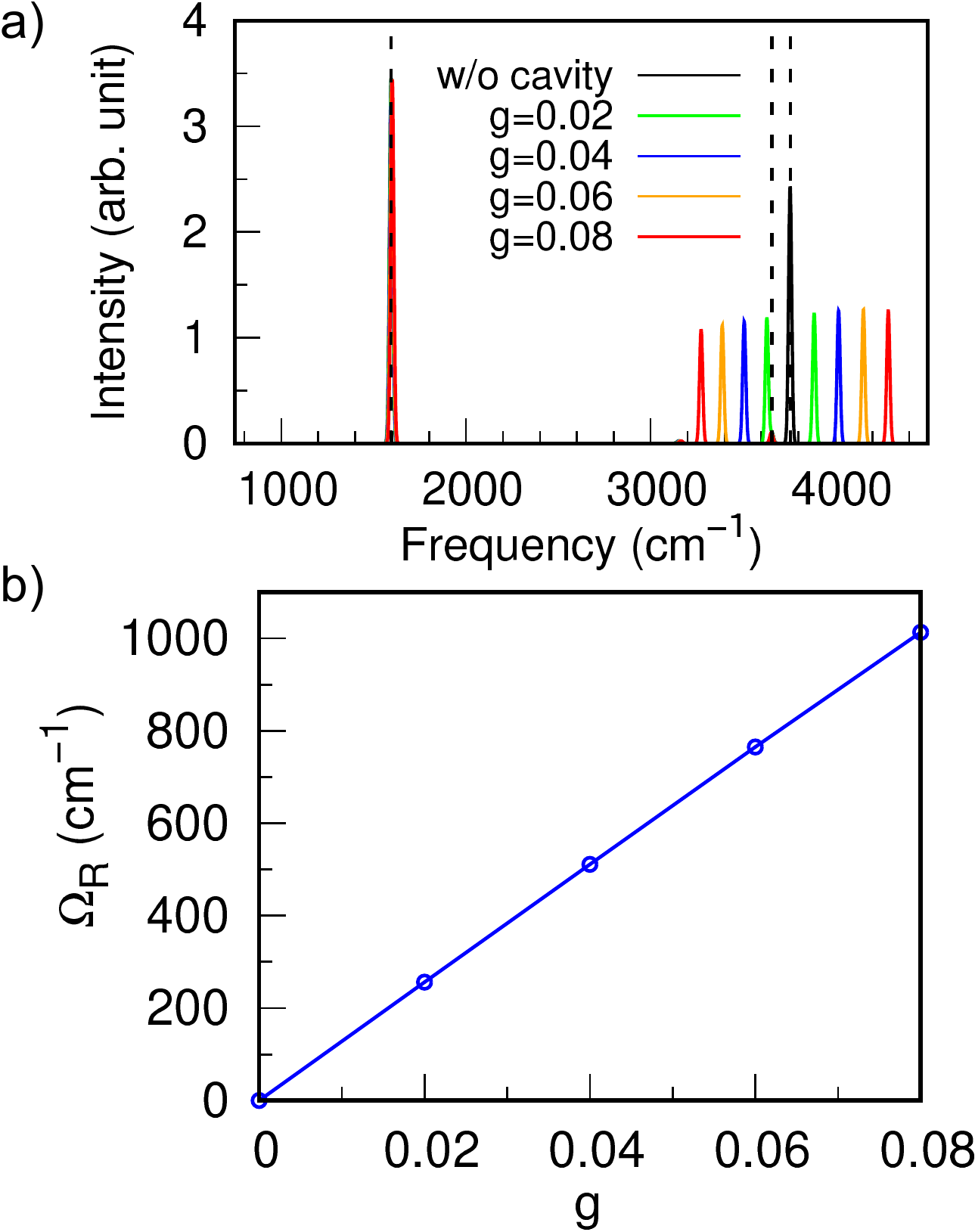}
\end{center}
\caption{(a) IR spectra of \ce{H2O} with different light-matter coupling factor g. The single cavity mode has frequency of 3756 cm$^{-1}$ polarized in y axis. Vertical dashed lines are experimental frequencies outside the cavity.\cite{Shimanouchi} (b) Rabi splitting frequencies $\Omega_{\text{R}}$ with different values of light-matter coupling factor g.  }
\end{figure}

Figure 1(c), 1(d) and 1(e) show the effective potential energy surface, as defined by Equation (8), along cavity mode and different modes of \ce{H2O} molecule, both within and outside the cavity. For Figure 1(c) where the single cavity mode is polarized along x axis, the surface of cavity-bending mode (left) is distorted significantly comparing with the corresponding surface in Figure 1(e) where the molecule is outside the cavity. This is simply because the polarization direction of the cavity mode is collinear with the 
dipole derivative direction (x axis) of the molecule. Thus the coupling between cavity mode and \ce{H2O}'s bending motion is maximized. The potential energy surface on cavity-symmetric stretch changes slightly for the molecule inside or outside the cavity, due to the minor change of x-component dipole along symmetric stretch and small coupling with the cavity mode. As to the surface of cavity-asymmetric stretch (right), it keeps almost the same comparing with the case outside the cavity (right panel of Figure 1(e)) since almost no change of the x-component dipole for the asymmetric stretch. When the cavity mode is polarized in y axis, the coupling behavior between the cavity mode and molecule is totally different. As seen in Figure 1(d), no changes are seen for the surfaces of cavity-bending and cavity-symmetric stretch. Instead, the surface of cavity-asymmetric stretch mode is distorted significantly, relative to the right panel of Figure 1(e). Again, this results from the fact that the z-component of dipole undergoes significant change along the asymmetric stretch and strongly couples with the cavity mode. 

Next, we move to the calculated cVSCF/VCI spectra of single water molecule in an optical cavity by varying the cavity mode frequencies, polarization directions and also the light-matter coupling factor g. Start from the system of single cavity mode with polarization direction along y axis and frequency of 3756 cm$^{-1}$, Figure 2(a) shows how the IR spectra of system changes with different values of light-matter coupling factor. When the molecule is outside the cavity, there are two intense peaks at 1594 and 3756 cm$^{-1}$. They correspond to the fundamental bands of water bending and asymmetric stretch modes respectively. There also exists a band at 3656 cm$^{-1}$ with small IR intensity which is the fundamental band of the symmetric stretch. Both peak positions and intensities agree excellently with the experimental data for the \ce{H2O}\cite{Shimanouchi,Burcl2003} which verifies the robustness of our VSCF/VCI approach. When the \ce{H2O} molecule is in the cavity and the cavity mode is polarized in y axis, the original asymmetric stretch band is splitted to two intense peaks while nothing changes for the bending and symmetric stretch bands. This agrees with the observation in Figure 1 that only asymmetric mode can couple with the y-axis polarized cavity mode. Moreover, as expected, the splitting increases with larger light-matter coupling factor, g. In Figure 2(b), we show the relationship between Rabi splitting frequencies and the coupling factor g. When g is great than 0.06, the corresponding coupling strength $\eta$ calculated from $\Omega_R/(2\omega_0)$ is larger than 0.1. Thus, the molecule is under vibrational ultrastrong coupling regime with g $\ge 0.06$. Note that in the experimental environment, the real light-matter coupling strength is a collective property which is proportional to $\sqrt{N_{\text{mol}}}$ where $N_{\text{mol}}$ is the number of molecules in a cavity. Here, we use a much larger coupling factor g just to compensate for the lack of large amount of molecules in this single-molecule calculation. The investigation of coupling factor g and setting up more realistic system are subject to future investigations.

\begin{table}[htbp!]
\centering
\caption{cVSCF/VCI state energies and leading VCI coefficient(s) under different values of light-matter coupling factor g and cavity mode polarized at x axis.}
\begin{threeparttable}
\begin{tabular*}{\columnwidth}{l | c c | c c | c c | c c |c c | c c}
\hline
\hline
\multicolumn{11}{c}{Cavity mode frequency $\omega=3756$ cm$^{-1}$} &\\
\hline
Coupling factor g & \multicolumn{2}{c|}{0.00} & \multicolumn{2}{c|}{0.02} & \multicolumn{2}{c|}{0.04} & \multicolumn{2}{c|}{0.06} & \multicolumn{2}{c|}{0.08}\\
\hline
Energy (cm$^{-1}$) & 1594 & 3756$^a$ & 1588 & 3777 & 1570 & 3832 & 1543 & 3913 & 1508 & 4017 \\
VCI coeff (bend) & 1.00 & / & 0.99 & 0.10 & 0.97 & 0.20 & 0.93 & 0.28 & 0.88 & 0.36\\
VCI coeff (sym-str) & / & / & / & 0.22 & / & 0.29 & / & 0.30 & / & 0.31\\
VCI coeff (cavity) & / & 1.0 & 0.11 & 0.97 & 0.21 & 0.93 & 0.29 & 0.89 & 0.36 & 0.86\\
\hline
\multicolumn{11}{c}{Cavity mode frequency $\omega=1594$ cm$^{-1}$} &\\
\hline
Coupling factor g & \multicolumn{2}{c|}{0.00} & \multicolumn{2}{c|}{0.02} & \multicolumn{2}{c|}{0.04} & \multicolumn{2}{c|}{0.06} & \multicolumn{2}{c|}{0.08}\\
\hline
Energy (cm$^{-1}$) & 1594 & 1594$^a$ & 1377 & 1845 & 1196 & 2126 & 1046 & 2433 & 927 & 2749 \\
VCI coeff (bend) & 1.00 & / & 0.65 & 0.75 & 0.58 & 0.78 & 0.52 & 0.81 & 0.46 & 0.80\\
VCI coeff (sym-str) & / & / & / & / & / & / & / & / & / & /\\
VCI coeff (cavity) & / & 1.0 & 0.75 & 0.65 & 0.78 & 0.59 & 0.79 & 0.54 & 0.78 & 0.50\\
\hline
\hline
\end{tabular*}
\begin{tablenotes}
\item[a] fundamental frequency of the cavity mode
\end{tablenotes}
\end{threeparttable}
\end{table}

\begin{figure}[htbp!]
\begin{center}
\includegraphics[width=0.45 \textwidth]{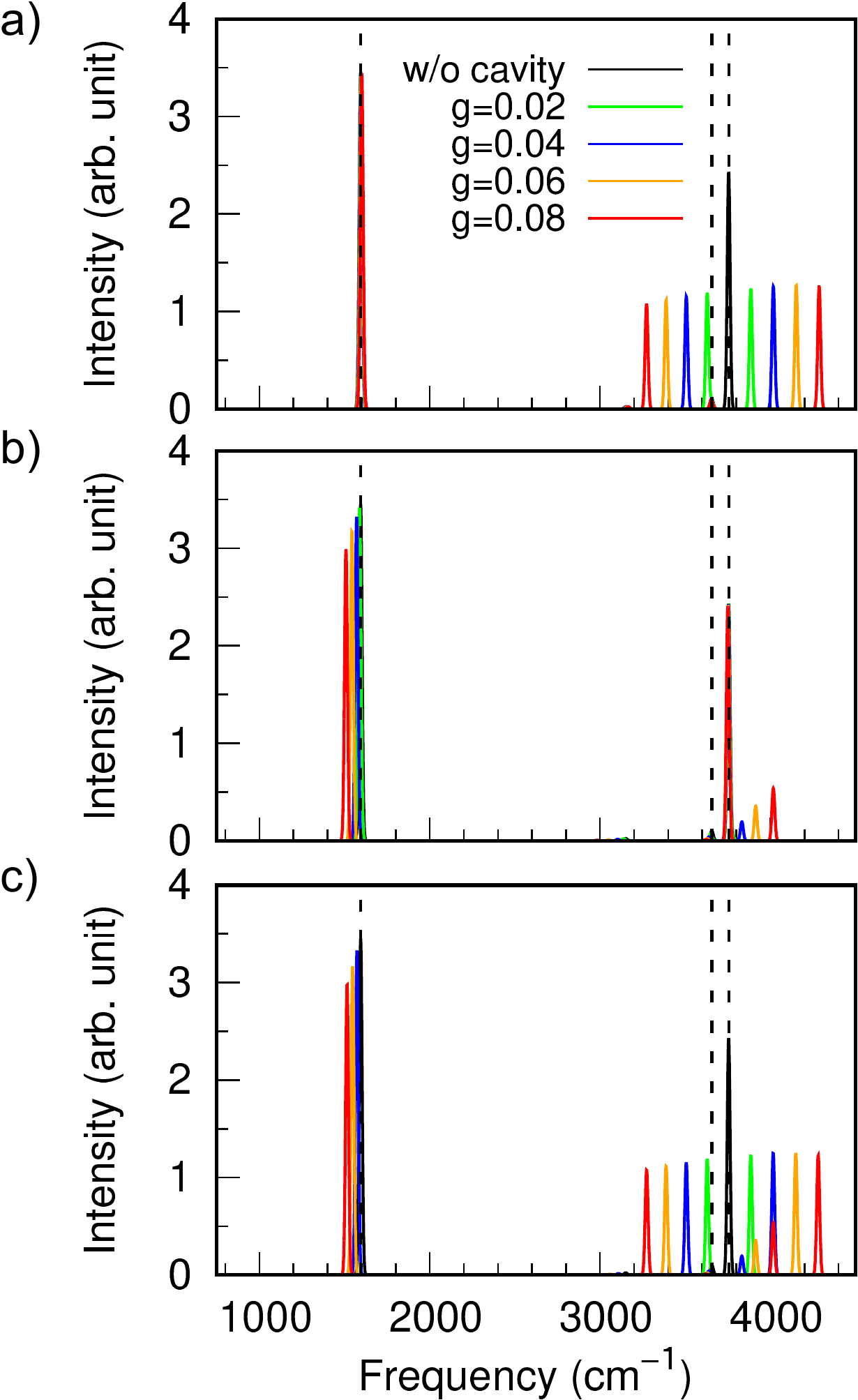}
\end{center}
\caption{IR spectra of \ce{H2O} with different values of light-matter coupling factor g for different cavity systems. (a) Single cavity mode with polarization direction along y axis, (b) Single cavity mode with polarization direction along x axis, (c) Two cavity modes with polarization direction along x and y axes respectively. The cavity mode has frequency of 3756 cm$^{-1}$ for all cases. Vertical dashed lines are experimental results outside the cavity.\cite{Shimanouchi}  }
\end{figure}

Figure 3 shows the IR spectra with different polarization directions of the cavity mode. When the single cavity mode is polarized in x axis (see Figure 3(b)), there does not exist splittings for the asymmetric stretch band because there is almost no coupling with the cavity for this mode. Instead, it is observed that the band for bending mode undergoes red shift with larger coupling strength. A new peak appears above 3756 cm$^{-1}$ and undergoes significant blue shift for large coupling strength. This high-frequency band indicates the formation of upper polariton (UP) while the red-shifting peak around 1594 cm$^{-1}$ corresponds to the signature of the lower polarition (LP).The UP state is attributed to the coupling between cavity mode, water bending motion and also the water symmetric stretch mode. Thus this hybrid state carries the signatures from both water bending and symmetric stretch and becomes IR active with significant intensity. In the upper part of Table 1, we list the frequency of this polariton state along with the leading VCI coefficients of different bases. For example, when $g=0.04$ a.u., the UP polariton state locates at 3832 cm$^{-1}$, higher than the initial cavity mode frequency of 3756 cm$^{-1}$. This hybrid state has contributions from the bending, symmetric stretch and also the cavity mode with corresponding CI coefficients 0.20, 0.29 and 0.93 respectively. It should be noted that, when the molecule is outside the cavity, we cannot observe a state where the bending and symmetric stretch are directly mixed. In the cavity, the cavity mode induces the direct mode coupling between these two important motions. Besides, with larger light-matter coupling strength, the mixing between water bend and symmetric stretch becomes stronger. This indicates that the existence of optical cavity, in both VSC and V-USC regions, can induce the coupling between molecule's motion and may greatly accelerate the intramolecular vibrational energy transfer. In Figure 3(c), we show the spectra when two cavity modes are included with the same frequency but different polarization directions. Comparing with Figure 3(a) and 3(b), this spectrum is almost a combination of the spectra of two single cavity mode systems, where we see the splittings of the 3756 cm$^{-1}$ asymmetric stretch band, the red-shifting peak below 1594 cm$^{-1}$ and the blue-shifting peak above 3756 cm$^{-1}$.

We also modified the frequency of cavity mode to 1594 cm$^{-1}$ to enable direct resonance with the water bending mode. When the polarization direction of the single cavity mode is set along x axis, the calculated cVSCF/VCI spectrum is shown in Figure 4(a). As seen, the bending mode undergoes splittings upon the molecule is within the cavity. The spectral positions of the symmetric and asymmetric stretches change by about 10 cm$^{-1}$ due to the slightly distorted potential energy surface and the mode couplings with complicated combination bands that involve the cavity mode. Similar to Figure 2(b), Figure 4(b) shows the relationship between Rabi splitting frequency associated with the bending motion and the light-matter coupling factor g. The light-matter coupling strength, $\eta$, is larger than 0.1 when g = 0.02. However, unlike Figure 3(b) with cavity frequency of 3756 cm$^{-1}$, in all the V-USC cases, we do not observe direct coupling between water bending and symmetric stretch for all the polariton states. Detailed cVSCF/VCI state energies and leading VCI coefficients are also listed in Table 1. It can be seen that when the molecule is within the cavity, only bending mode is involved in the generation of both LP and UP states. This is mainly because the cavity mode frequency is far off the symmetric stretch and their direct coupling is almost negligible. In Figure 5(b), we show the cVSCF/VCI spectra of \ce{H2O} molecule in the cavity where single cavity mode is included with frequency of 1594 cm$^{-1}$ and polarized in y axis. In this case, the cavity mode does not have direct resonance with the water bending mode just because the motion of bending mode results in dipole moment change only along x direction. Instead, the cavity mode couples significantly with the asymmetric stretch of water molecule and results in large blue shift of the asymmetric stretch band around 3756 cm$^{-1}$. These high-frequency bands corresponds to the UP states and the LP states also appears in the region below 1594 cm$^{-1}$ with very small IR intensity. The low IR intensities of these LP states is because the VCI coefficient of asymmetic stretch is very small and the LP is still dominated by the cavity mode.

\begin{figure}[htbp!]
\begin{center}
\includegraphics[width=0.46\textwidth]{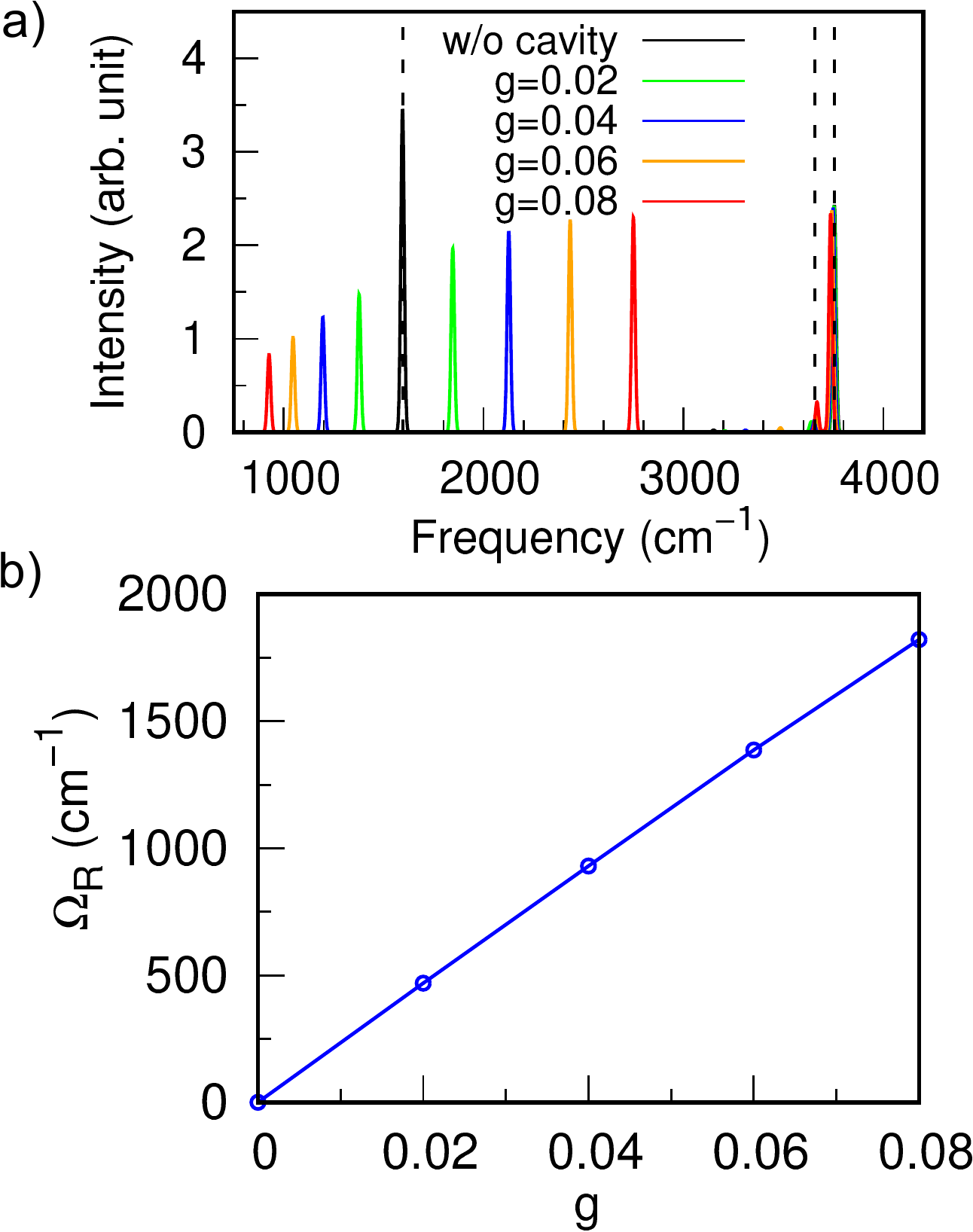}
\end{center}
\caption{(a) IR spectra of \ce{H2O} with different values of light-matter coupling strength g. The single cavity mode has frequency of 1594 cm$^{-1}$ with polarization direction along x axis. Vertical dashed lines are experimental frequencies without cavity.\cite{Shimanouchi} (b) Rabi splitting frequencies $\Omega_{\text{R}}$ with different values of light-matter coupling factor g.  }
\end{figure}

Finally, We considered two cavity modes with polarization direction in x and y axes separately and frequency of 1594 cm$^{-1}$ and the calculated IR spectrum is shown in Figure 5(c). When the light-matter coupling factor g changes, similar to the previous conclusion in Figure 3(c), the IR spectrum with two cavity modes is almost a combination of the two spectra of single cavity modes (Figure 5(a) and 5(b)). However, when the light-matter coupling factor g equals to 0.06 in the ultra strong vibrational coupling regime, besides of the usual splitting of the water bending signature, an additional splitting pattern appears around 3900 cm$^{-1}$. This surprising feature indicates additional vibrational resonance between water asymmetric stretch and cavity modes. By analyzing the VCI coefficients of these two states, it is found that both of them involves significant contributions from the water bending motion which is not observed in other cases. Actually, this can be viewed as a ``two-step'' process. An upper polariton state at 2433 cm$^{-1}$ is generated due to the direct resonance between the water bending mode and the x-axis polarized cavity mode. At the same time, another upper polariton state should be generated in principle at 3960 cm$^{-1}$ (see Figure 5(b)) due to the coupling between the water asymmetric stretch and the y-axis polarized cavity mode. However, since the y-axis polarized cavity mode is included, a combination band between the UP state at 2433 cm$^{-1}$ and the y-axis polarized cavity mode (1594 cm$^{-1}$) locates just around 3960 cm$^{-1}$. This directly results in the resonance between the new combination bands and the UP state at 3960 cm$^{-1}$. Thus, the original 3960 cm$^{-1}$ band is split into two states and both of them include the vibrational components of water bending and asymmetric stretch modes. Detailed illustrative scheme of of how the new splitting pattern is generated is shown in Figure S3 in SI. To the best of our knowledge, the direct strong coupling between water bending and asymmetric stretch fundamental bands has been observed previously both theoretically and experimentally. When the water molecule is included in the optical cavity with ultra strong vibrational coupling conditions, their directing coupling can be realized with cavity modes as intermediary. This also indicates that the intramolecular vibrational energy transfer between the water bending and asymmetric stretch modes can be greatly improved with the existence of the optical cavity. For a realistic optical cavity, many cavity modes are need to be considered.\cite{Li2020_pnas,Ribeiro2021} In Figure S4 of the SI, we show the preliminary results of the IR spectra with more cavity modes included. It is observed that more intense spectral peaks appears in the spectrum range of [0,6000] cm$^{-1}$ which resulting from complicated coupling between water and cavity modes. These preliminary simulations of multiple cavity mode systems also verify the fact that the direct and strong coupling between water bending, symmetric and asymmetric stretches can be induced by the cavity system under either VSC or V-USC conditions.

\begin{figure}[htbp!]
\begin{center}
\includegraphics[width=0.45\textwidth]{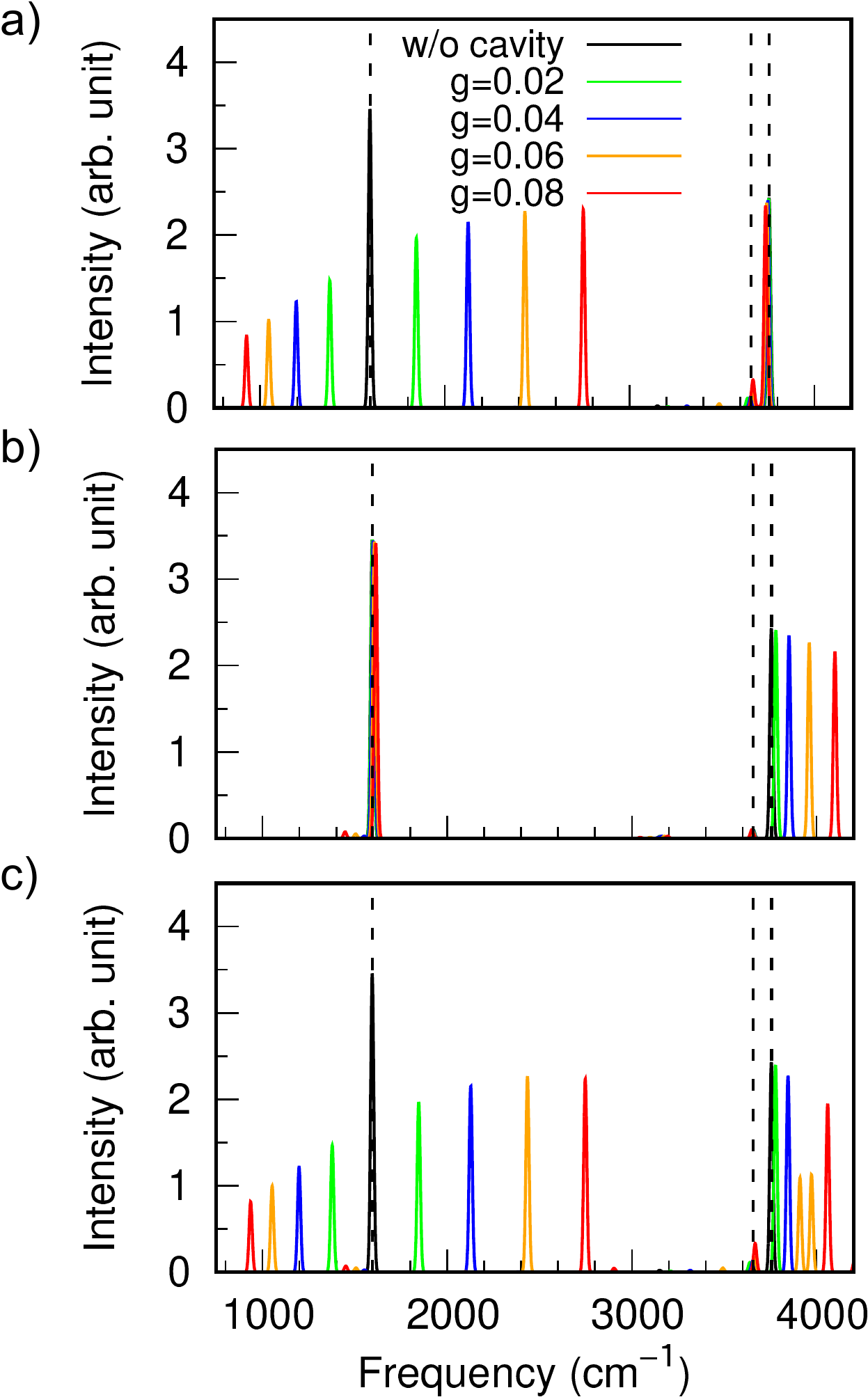}
\end{center}
\caption{IR spectra of \ce{H2O} with different values of light-matter coupling strength g for different cavity systems. (a) Single cavity mode with polarization direction along x axis,
(b) Single cavity mode with polarization direction along y axis,
(c) Two cavity modes with polarization direction along x and y axes respectively. The cavity mode has frequency of 1594 cm$^{-1}$ for all cases. Vertical dashed lines are experimental results without cavity.\cite{Shimanouchi}  }
\end{figure}
\newpage
To summarize, we have presented the cavity-VSCF/VCI approach which enables fully quantum vibrational analyses of the molecular system under VSC or V-USC. The robustness of cVSCF/VCI approach and accuracy of the potential energy/dipole moment surfaces enable quantitative analyses of the virbational spectra of molecular system in an optical cavity. Rabi splittings are observed in the IR spectra when the cavity mode had resonance with certain vibrational mode of water. The spectral signatures in the splitting region and the blue/red shift of certain bands are highly related with the frequency and polarization directions of the cavity mode. Further analyses of the vibrational bands in the IR spectra found the unusual fact in terms of vibrational mode couplings between water modes. Through modifications of the number of cavity modes, frequencies and polarization directions, the water bending mode can be strongly coupled with the symmetric and asymmetric stretches which can not be seen when the molecule is outside the cavity. This opens the new possibilities of realizing fast intramolculear vibratinal energy transfer under VSC or V-USC conditions within an optical cavity. 

The current work focuses on single molecule within a cavity. However, as mentioned, in the realistic cavity systems, the light-matter coupling is a collective behavior which involves large amount of molecules. To conduct fully quantum vibrational analyses of the realistic cavity system, the current light-matter coupling factor g should be rescaled to a much smaller value and the cVSCF/VCI approach should also be further modified with certain approximations, such as decreasing the level of mode couplings and reducing the dimensionality by choosing partial amount of normal modes. For simple molecular systems like \ce{H2O}, the IR spectra can be approximated through simplified quantum Hamiltonian models. However, for more complicated and strongly anharmonic systems, such as hydrated water system, advanced quantum methods are need to reproduce the experimental measurements and provide quantitative explanations. The theoretical model presented in this letter provides a practical and accurate way to investigate how polariton state are generated and how molecular's vibrational motions are affected by the cavity. The calculated IR spectra also indicates new possible pathway to realize the intramolecular vibrational energy transfer by adjusting several key parameters of the cavity systems. Such process can be further verified through time-dependent dynamics simulations which is subject to our future directions. The cVSCF/VCI approach can also be applied to simulate 2D-IR spectra which helps probe the special features of the potential and understand the vibrational dynamics.\cite{Yu2020,Yu2021}This work can be taken as benchmark for other approximate methods and provides the foundations for a wide range of theoretical and experimental investigations in the future. 

\section*{Acknowledgment}
QY thanks Dr. Tao E. Li, Professor Raphael Ribeiro and Professor Joel Bowman for helpful discussions and suggestions. QY thanks Professor Sharon Hammes-Schiffer and National Science Foundation (Grant No. CHE-1954348) for support.

\section*{Supporting Information available}
- N-mode representation of the potential in VSCF/VCI calculation;\\
- Additional vibrational spectral with differenty set up of the cavity system;\\
- Detailed scheme of formation of polariton states and the new splitting feature in Figure 5.c.

\bibliography{refs}

\end{document}